\documentclass[referee]{raa}           
\usepackage{graphicx,times}
\usepackage{natbib}
\usepackage{amssymb,amsmath}
\bibpunct{(}{)}{;}{a}{}{,}

\usepackage{hyperref}
\hypersetup{pdftitle = The title of my PDF, pdfauthor = My name, pdfsubject= The subject, pdfkeywords = keyword1 keyword2 keyword3} 
\hypersetup{colorlinks = true, linkcolor = green, anchorcolor = red, citecolor = blue, filecolor = red, pagecolor = red, urlcolor = red}

\begin{document}

   \title{Intra-day variability of BL Lacertae from 2016 to 2018
%\,$^*$
%\footnotetext{$*$ Supported by the National Natural Science Foundation of China.}
}
%   \subtitle{I. Place Your Subtitle Here}

   \volnopage{Vol.0 (20xx) No.0, 000--000}      %%preserved for Editor. DOn't remove!
   \setcounter{page}{1}          %%starting page, preserved for Editor. DOn't remove!

   \author{Tian Li
   \and Jianghua Wu
   \and Nankun Meng
   \and Yan Dai
   \and Xiaoyuan Zhang
   }
%% Here is an example of three authors come from different institutes.
%% For single author or all the authors from an institute, use "\inst{}" only

   \institute{Department of Astronomy, Beijing Normal University, Beijing 100875, China; {\it jhwu@bnu.edu.cn}\\
%% Please give the E-mail address of the author, to whom future correspondence and
%% offprint requests will be sent.
\vs\no
   {\small Received~~20xx month day; accepted~~20xx~~month day}}

\abstract{We monitored BL Lacertae in the B, V, R and I bands for 14 nights during the period of 2016-2018. The source showed significant intraday variability on 12 nights. We performed colour-magnitude analysis and found that the source exhibited bluer-when-brighter chromatism. This bluer-when-brighter behavior is at least partly caused by the larger variation amplitude at shorter wavelength. The variations at different wavelengths are well correlated and show no inter-band time lag.
}

   \authorrunning{Tian Li et al. }            %author_head in even pages
   \titlerunning{Intra-day variability of BL Lacertae}  % title_head in odd pages
   \maketitle

%________________________________________________ sections below
% 
\section{Introduction}           %% first-level sections will be auto-capitalized
\label{sect:intro}

Blazar is the most violently variable object among all kinds of active galactic nuclei (AGNs). The relativistic jets of blazars are believed to orient close to the line of our sight and powered by the central accretion disk-supermassive black hole systems. BL Lacertae object and flat-spectrum radio quasar are the subsets of blazar. The BL Lacertae object is named after the well-known blazar called BL Lacertae, which is characterized by its high and variable polarization, absence of strong emission lines in the optical spectrum, synchrotron emission from relativistic jets, and intense flux and spectral variability from radio to $\gamma$-ray on a wide variety of time-scales \citep{1995ARA&A..33..163W,2003ApJ...596..847B}.  For intraday variability (IDV), the flux can change over hundredth or even tenth magnitudes within several hours \citep{2015MNRAS.450..541A}. Blazar's spectral energy distribution (SED) displays two peaks \citep{1998MNRAS.299..433F}. We can divide the BL Lacs into 
three classes based on the locations of these peaks. For the high energy peaked BL Lacs (HBLs), their first peaks locate in UV/X-rays while the second peaks locate at TeV energies. The synchrotron emission peaks of Intermediate-frequency-peaked BL Lacs (IBL) lies in optical region. BL Lacertae is a low-frequency peaked BL Lacs (LBL) \citep{2004MNRAS.348.1379C,2010ApJ...716...30A} as its first component peaks at infrared while its second component peaks around MeV-GeV \citep{1995MNRAS.277.1477P,2010ApJ...716...30A}. 

BL Lacertae, hosted in a giant elliptical galaxy with R = 15.5 \citep{2000ApJ...544..258S}, has a redshift of z = 0.0668 $\pm$ 0.0002 \citep{1977ApJ...212L..47M}. It was once observed in several multiwavelength campaigns carried out by the Whole Earth Blazar Telescope (WEBT/GASP) \citep{2004A&A...424..497V,2006A&A...456..105B,2009A&A...507..769R}. Some other investigations have been carried out to study its flux variations, spectral changes, and inter-band cross-correlations \citep{1972ApJ...178L..51E,1992AJ....104...15C,2002A&A...390..407V,2003A&A...397..565P,2012A&A...538A.125Z,2015MNRAS.450..541A,2015MNRAS.452.4263G,2017MNRAS.469.3588M,2018Galax...6....2B,2020Galax...8...11S}. Most of the observations found its amplitude of IDV is larger at a shorter wavelength. The IDV amplitude is usually larger when the duration of the observation is longer \citep{2005A&A...440..855G, 2015MNRAS.452.4263G, 2017Galax...5...94G}. The IDV amplitude also decreases as the source flux increases \citep{2015MNRAS.452.4263G,2017Galax...5...94G}. The reason might be that the irregularities in the turbulent jet will decrease when the source is at a bright state and fewer non-axisymmetric bubbles were  carried downward in the relativistic jets \citep{2014ApJ...780...87M,2015MNRAS.452.4263G}. \citet{2018A&A...615A.118S} found a possible $\gamma$-ray and optical correlated quasi-periodicities of 1.86 yr. Bluer-when-brighter (BWB) trend was found in previous observations. The BWB trend tends to appear on short timescales rather than on long timescales, indicating that there are probably two different components in the variability of BL Laccertae \citep{2002A&A...390..407V,2004A&A...421..103V}. As BL Lac objects usually show BWB trends, redder-when-brighter trends are frequently seen in flat spectrum radio quasar (FSRQs) \citep{2018RAA....18..150L, 2019AJ....157...95G}. The variation in different bands are highly correlated. Several authors found time-lag between variations in different bands of BL Lacertae. For example, \citet{2003A&A...397..565P} found a delay of 0.4 h between the B and I bands. \citet{2006MNRAS.373..209H} found a delay of 11.6 minutes between the e and m bands. A possible time lag of 11.8 minutes between the R and V bands was reported by \citet{2017MNRAS.469.3588M}.

In this paper, we aim to study the optical IDV and spectral variations of BL Lac. We carried out photometric measurement of this object on 14 nights in 2016-2018. We also tried to find any possible inter-band time lags in order to study the physical nature of the acceleration and cooling mechanisms in the relativistic jet and the origin of IDV. The paper is organized as follows. In section 2, we describe our observations and data reductions. Section 3 describes the analysis techniques followed by results. Section 4 gives the conclusions.

\begin{table*}
	\centering
	\caption{Observation log}
	\label{log}
	\begin{tabular}{cccrrc} 
		\hline
		Julian Date & Date & Passbands & Data points&Exposure time & Duration\\
		&( yyyy mm dd )& & & (seconds) & (hours)\\
		\hline
		2457696 & 2016 11 03 & $V$ & 160 & 60 &6.2\\
		          &            & $R$ & 160 & 60 &6.2\\
		2457697 & 2016 11 04 & $V$ & 94 & 60 &3.6\\
		          &            & $R$ & 93 & 60 &3.6\\
		2457699 & 2016 11 06 & $V$ & 178 & 60 &7.0\\
		          &            & $R$ & 181 & 60 &7.0\\
		2457745 & 2016 12 22 & $B$ & 97 & 30 &2.3\\
		          &            & $V$ & 97 & 20 &2.3\\   
		          &            & $R$ & 95 & 8 &2.2\\  
		2457746 & 2016 12 23 & $B$ & 103 & 40&2.8\\
		          &            & $V$ & 101 & 18&2.8\\ 
		          &            & $R$ & 101 & 10&2.8\\ 
		2458012 & 2017 09 15 & $B$ & 232 & 60&8.3\\
		          &            & $R$ & 250 & 20&8.5\\
		          &            & $I$ & 233 & 20&8.3\\
		2458013 & 2017 09 16 & $B$ & 235 &60 &8.2\\
		          &            & $R$ & 235 &20 &8.4\\
		          &            & $I$ & 232 &20 &8.1\\
		2458014 & 2017 09 17 & $B$ & 230 & 60&8.3\\
		          &            & $R$ & 236 & 20&8.5\\
		          &            & $I$ & 228 & 20&8.3\\
		2458369 & 2018 09 07 & $B$ & 31 & 60&2.3\\
		          &            & $V$ & 31 & 60&2.3\\
		          &            & $R$ & 31 & 60&2.3\\
		          &            & $I$ & 31 & 60&2.3\\
		2458370 & 2018 09 08 & $B$ & 52 & 60&3.8\\
		          &            & $V$ & 52 & 60&3.8\\
		          &            & $R$ & 52 & 60&3.8\\
		          &            & $I$ & 52 & 60&3.8\\          
		2458420 & 2018 10 28 & $B$ & 67 & 60&5.1\\
		          &            & $V$ & 68 & 60&5.1\\
		          &            & $R$ & 68 & 60&5.0\\
		2458424 & 2018 11 01 & $R$ & 132 & 50&4.2\\
		          &            & $I$ & 132 & 40&4.2\\
		2458425 & 2018 11 02 & $B$ & 84 & 20&3.5\\
		          &            & $V$ & 84 & 40&3.5\\
		          &            & $R$ & 84 & 60&3.5\\
		2458456 & 2018 12 03 & $B$ & 51 & 120&4.3\\
		          &            & $V$ & 51 & 60&4.3\\
		          &            & $R$ & 51 & 60&4.2\\
		\hline
	\end{tabular}
\end{table*}

\section{Observations and data reductions}

The observations were carried out on 14 nights in the period from 2016 November 3 to 2018 December 3. We used an 85 cm reflector to do the observations. It is at Xinglong Station, National Astronomical Observatories, Chinese Academy of Science (NAOC). The telescope uses primary focus system (F/3.27) with an Andor CCD and Johnson and Cousins filters UBVRI. The CCD has 2048$\times$2048 pixels and the pixel size is 12 $\mu$m. 

\begin{figure}
	\centering
	\includegraphics[width=0.5\columnwidth]{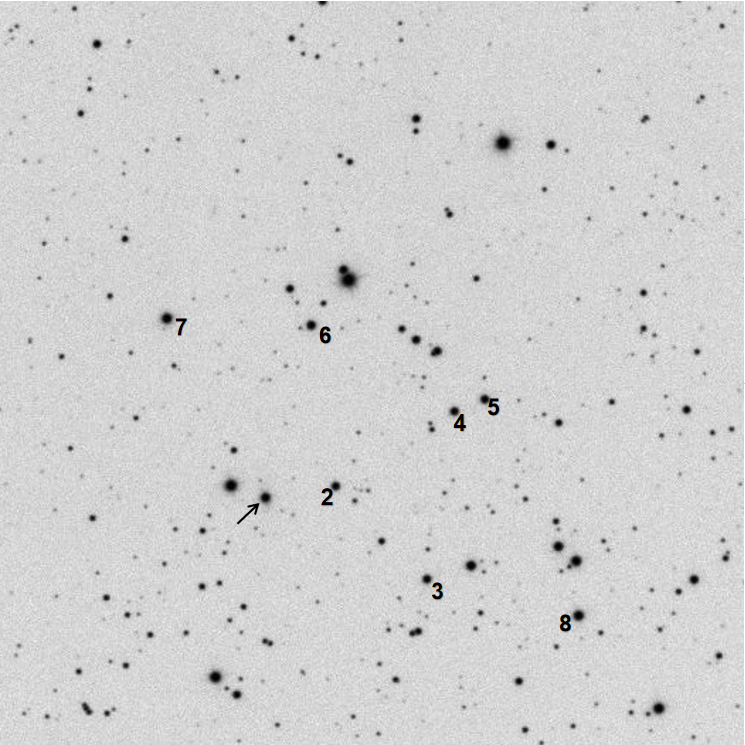}
    \caption{Finding chart of BL Lacertae in $R$ band. The images was taken on 2016 November 03. The BL Lacertae is marked by an arrow. Star 6 is the check star,  Star 2 and Star 4-8 were selected and used in the quantitative assessment of the IDV. Their magnitudes were all calibrated relative to the brightness of star 3.}
    \label{findchart}
\end{figure}

The photometric observations were performed in the B, V, R and I bands, we chose different combination of filters on different observations (see Table \ref{log}). The camera was switched to a cyclical mode for the exposures. In order to get enough signal to noise ratio (SNR), the exposure times were set according to the filter, weather condition, seeing, moon phase, and atmospheric transparency. It ranges from 8 to 120 seconds. The observation log is shown in the Table \ref{log}. Figure \ref{findchart} shows the finding chart.

We used the IRAF to reduce the data. The procedures included bias subtraction, flat fielding, extraction of instrumental aperture magnitude, and flux calibration. The average FWHM of the stellar images varied between 2 and 4 arcsecs from night to night. After a few trials with different aperture sizes, we adopted an aperture size of 1.5 times of the average Full width at half maximum (FWHM) of the stellar images. The inner and outer radii of the sky annuli were adopted as 5 and 7 times of the stellar FWHM, respectively. The magnitudes of BL Lacertae were calibrated with respect to the magnitude of star 3 in Figure \ref{findchart}. star 6 is selected as the check star. Its magnitudes were also calibrated and were used to check the accuracy of our observations. Star 2 and stars 4-8 were selected and used in the quantitative assessment of the IDV, as will be described in the next section. Their magnitudes were calibrated relative to the brightness of star 3. The standard magnitudes of stars 3, 4, and 6 in the B, V, R and I bands are given by \citet{1985AJ.....90.1184S}.

The photometric errors from IRAF are significantly underestimated according to \citet{2013JApA...34..273G}. Their method is to determine a coefficient $\eta$ which is the ratio between the real photometric error and that given by IRAF. Here we selected the check star (star 6) due to its lowest fluctuations and calculate $\chi$ by using the equation
\begin{equation}\label{chi2}
\chi^{2}=\displaystyle\sum_{i=1}^{N}\frac{(V_i-\overline{V})^2}{\sigma_{i}^2}.
\end{equation}
In this equation, $V_i$ is the ith differential magnitude, $\overline{V}$ is the mean of all differential magnitudes and $\sigma_{i}$ is the original error given by IRAF. The degree of freedom $\nu$ can be calculated from:
\begin{equation}
\nu=N-1=\chi^2/\eta^2.
\end{equation}
Then we obtained the regression analysis with fixed slope to calculate the coefficient $\eta$. See \citet{2013JApA...34..273G} for more details.

We calculated the $\eta$ of each band in each day and find that the $\eta$ ranges from 1.0764 to 1.7367, which is used to modify the original errors obtained by IRAF.

\section{Results}
\begin{figure*}
    \centering
	\includegraphics[width=\textwidth]{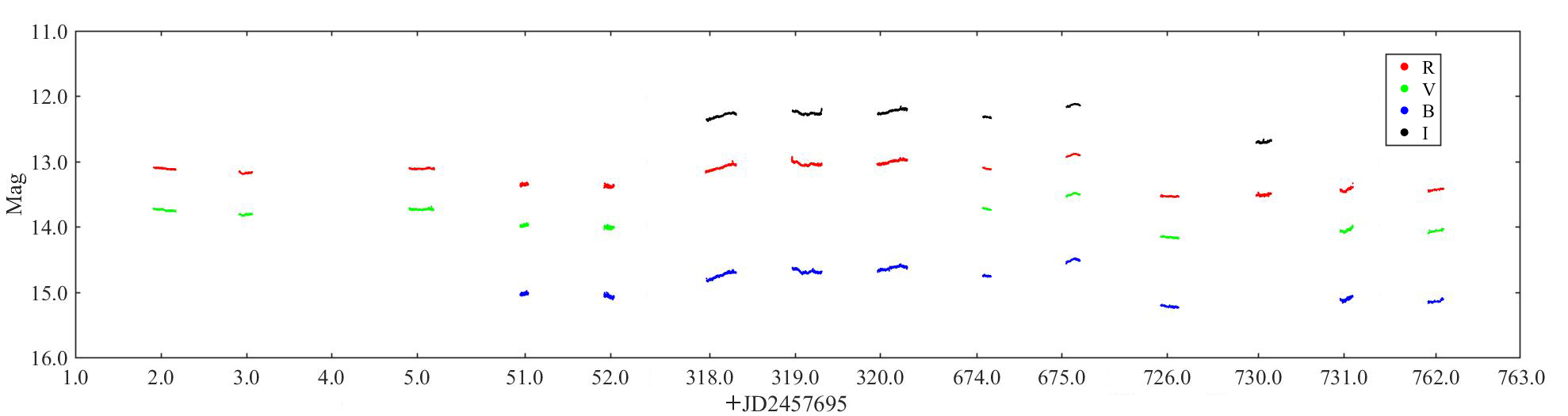}
    \caption{Light curves of BL Lacertae in the B, V, R and I bands in 3 years. Different colour dots represent data in different bands.}
    \label{overall}
\end{figure*}
\subsection{Light curves}
\label{sec:Lightcurve}

The overall light curves are displayed in Figure \ref{overall}. The light curves show obvious long term fluctuations. The largest amplitude of R band is about 0.6 mag.

\begin{figure*}
    \centering
    \begin{tabular}{ccc}
	\includegraphics[width=0.33\textwidth]{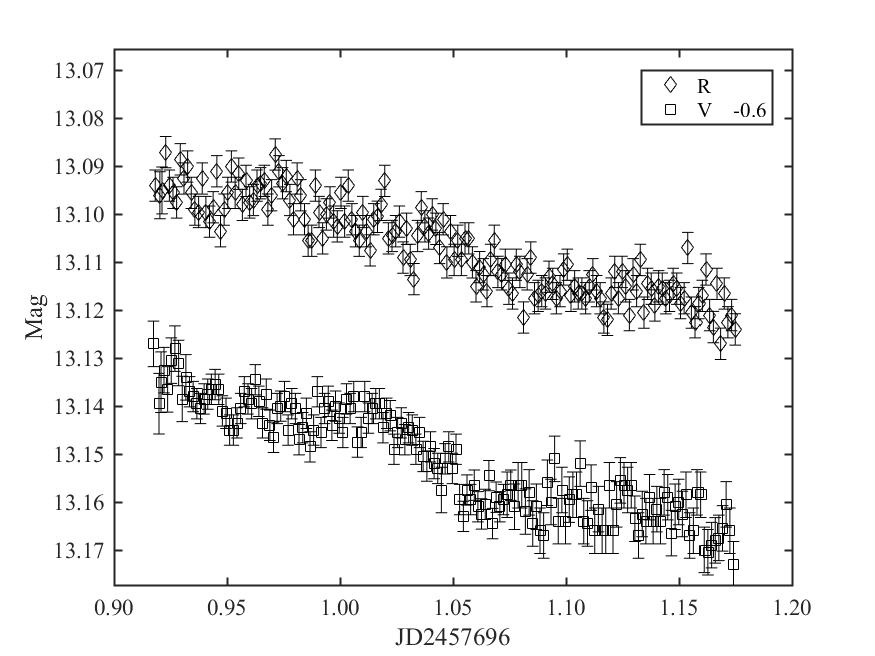}
	\includegraphics[width=0.33\textwidth]{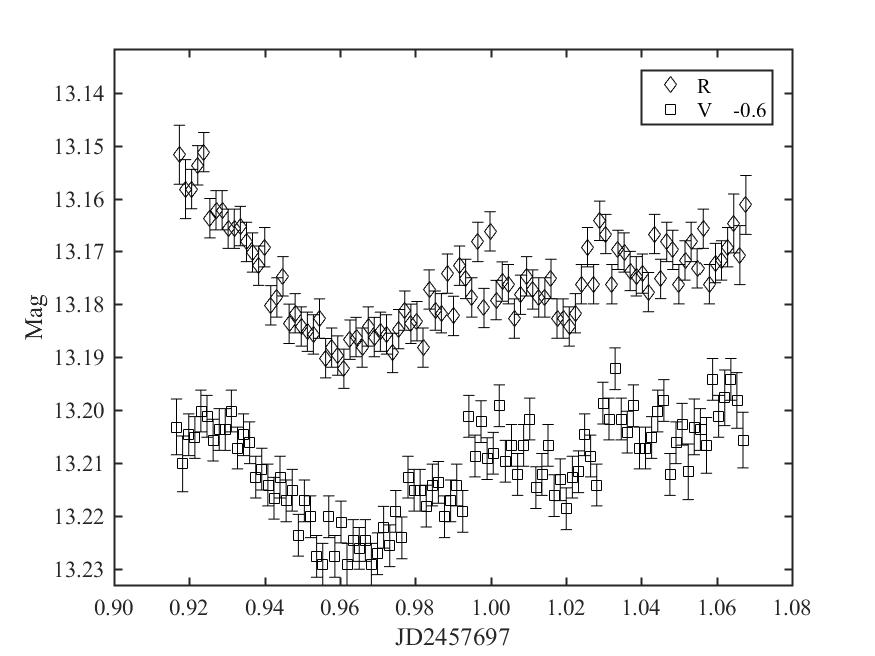}
	\includegraphics[width=0.33\textwidth]{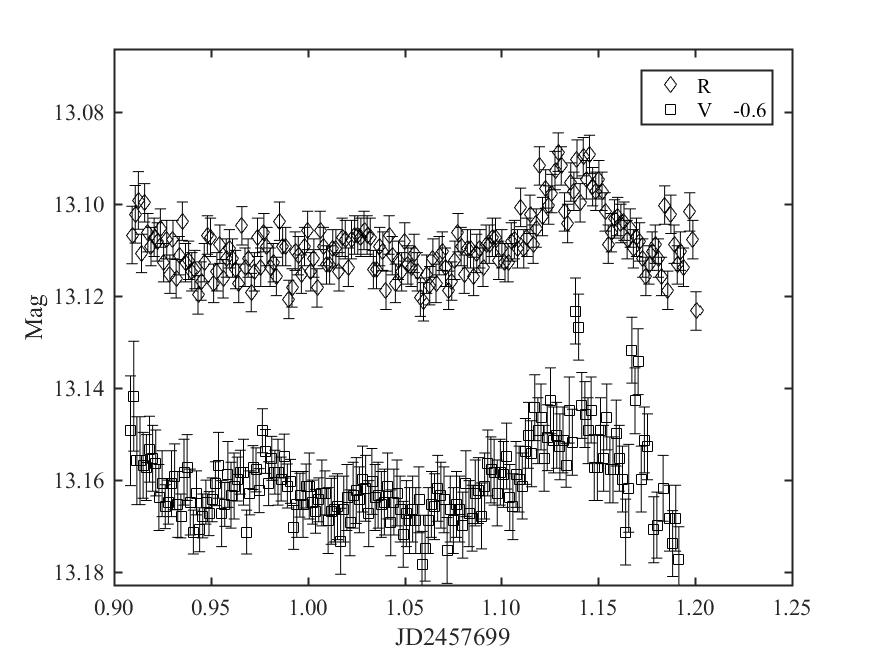}\\
	\includegraphics[width=0.33\textwidth]{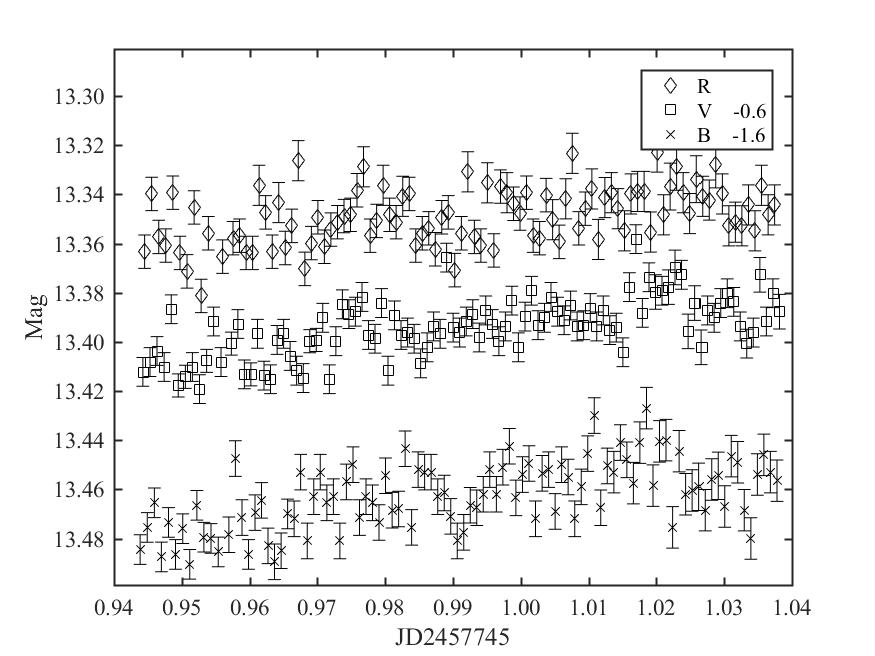}
	\includegraphics[width=0.33\textwidth]{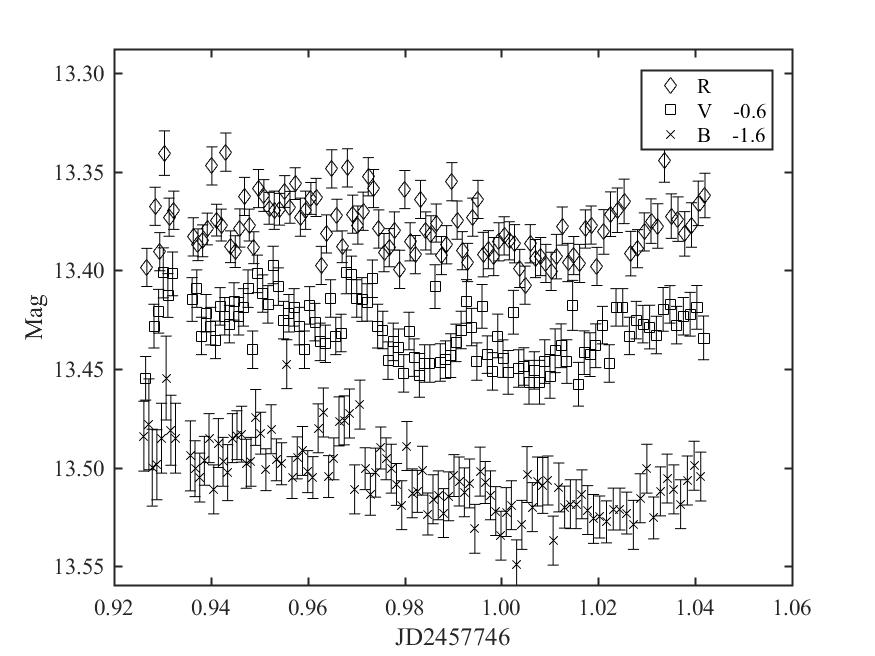}
	\includegraphics[width=0.33\textwidth]{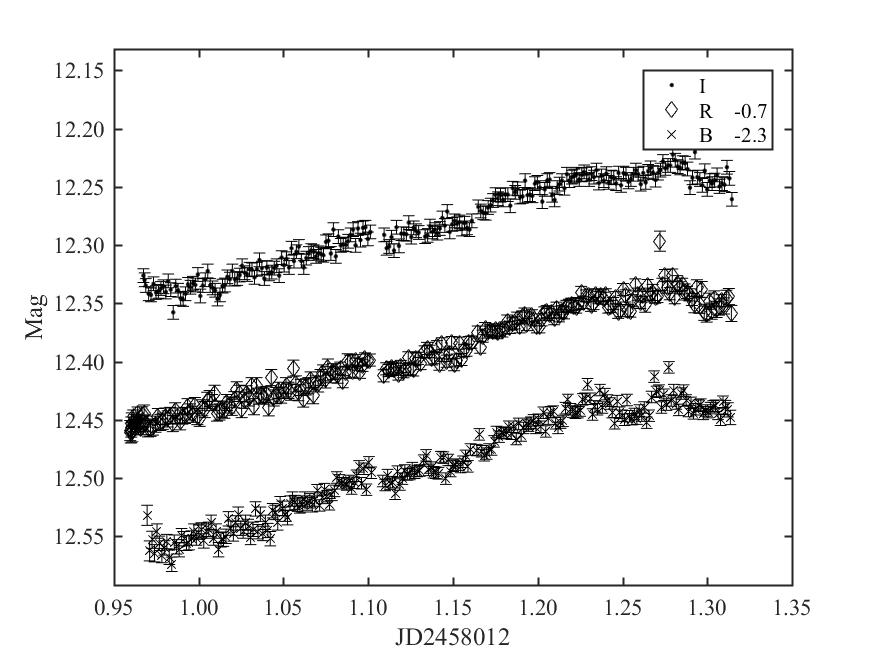}\\
	\includegraphics[width=0.33\textwidth]{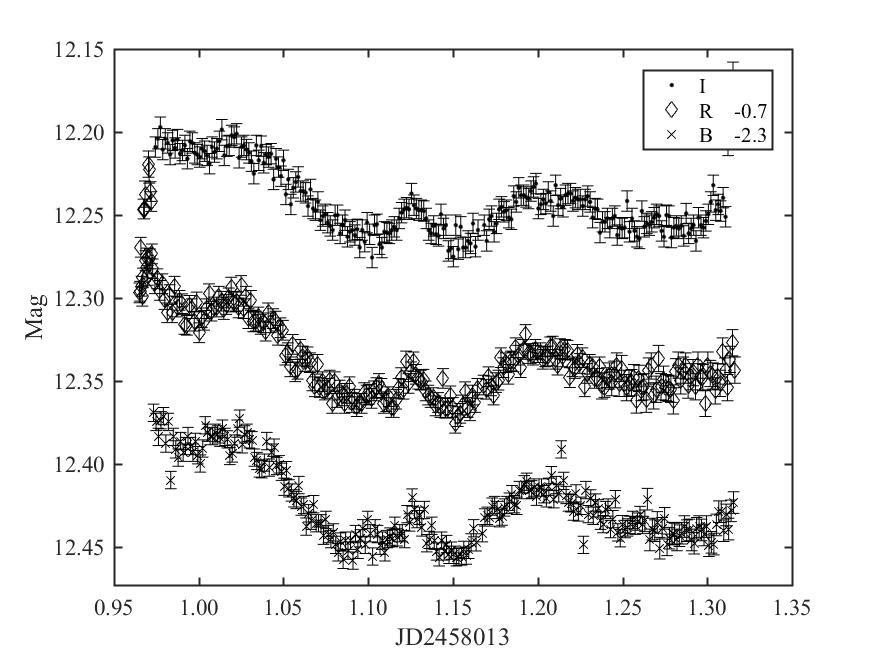}
	\includegraphics[width=0.33\textwidth]{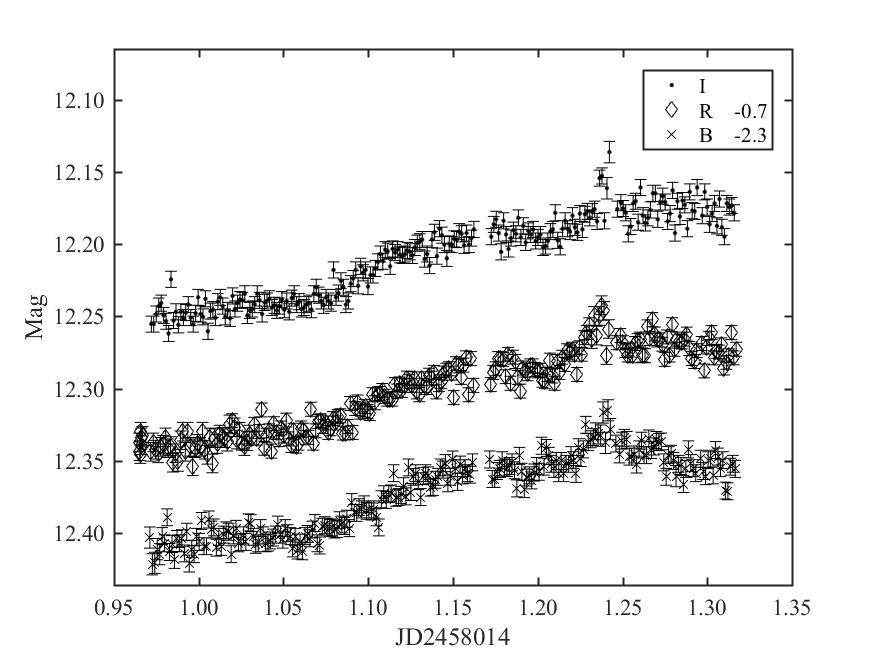}
	\includegraphics[width=0.33\textwidth]{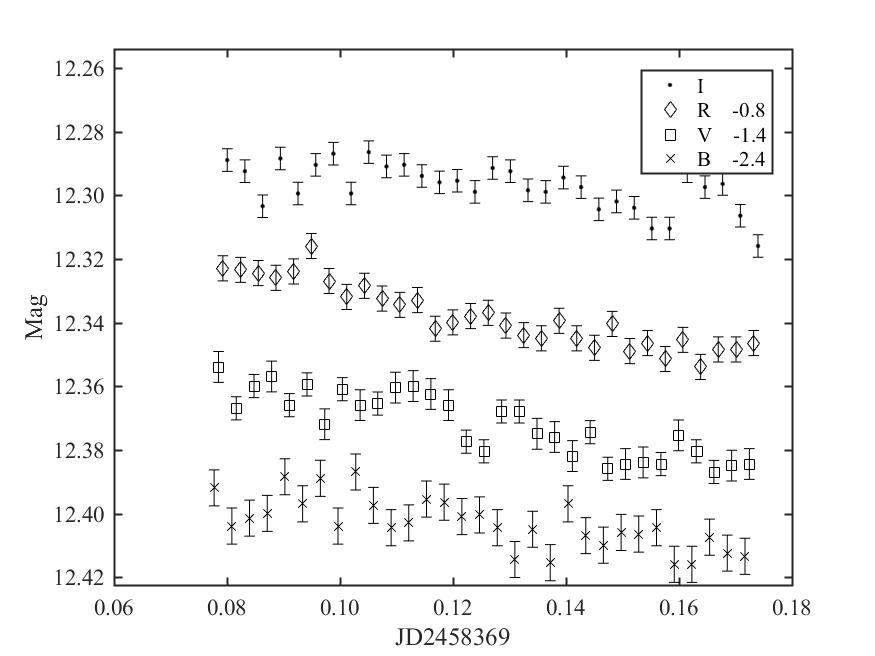}\\
	\includegraphics[width=0.33\textwidth]{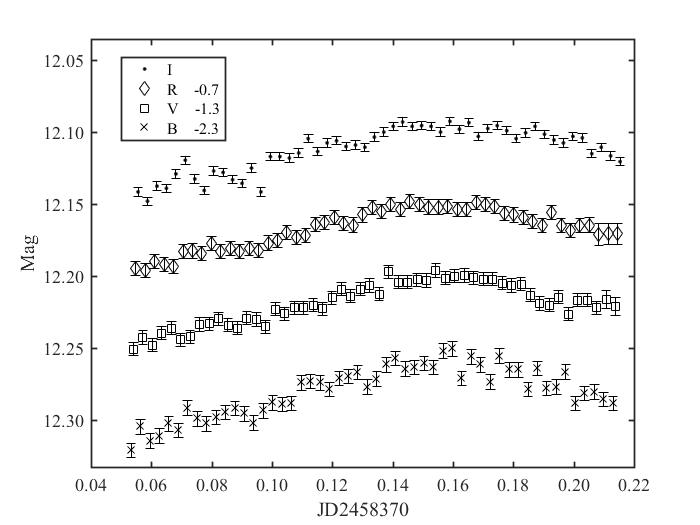}
	\includegraphics[width=0.33\textwidth]{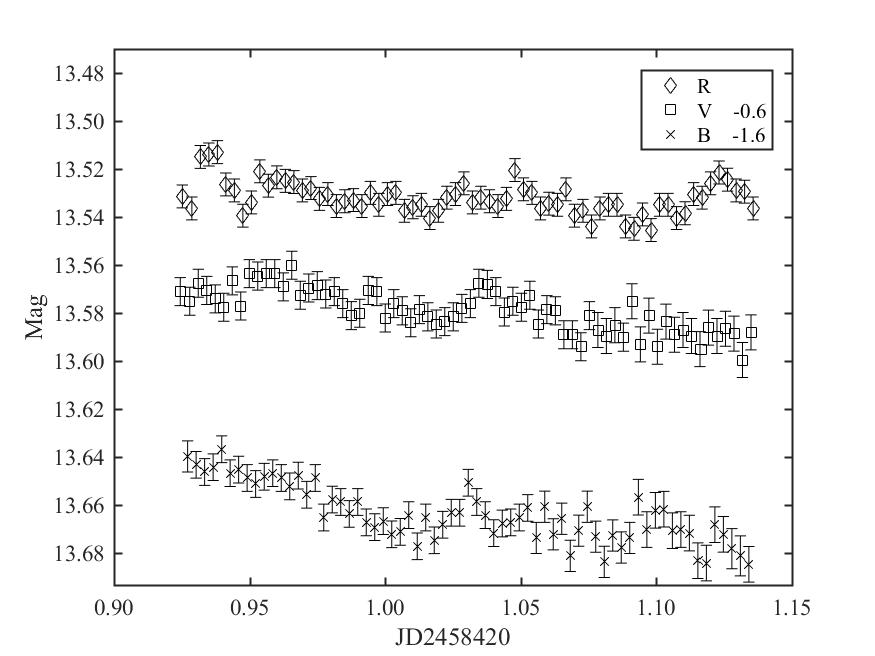}
	\includegraphics[width=0.33\textwidth]{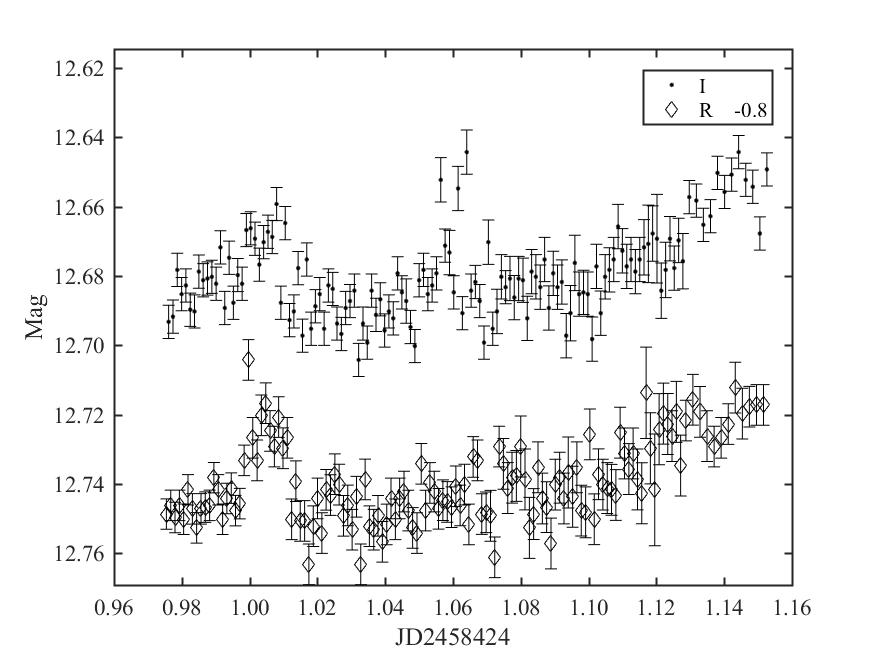}\\
	\includegraphics[width=0.33\textwidth]{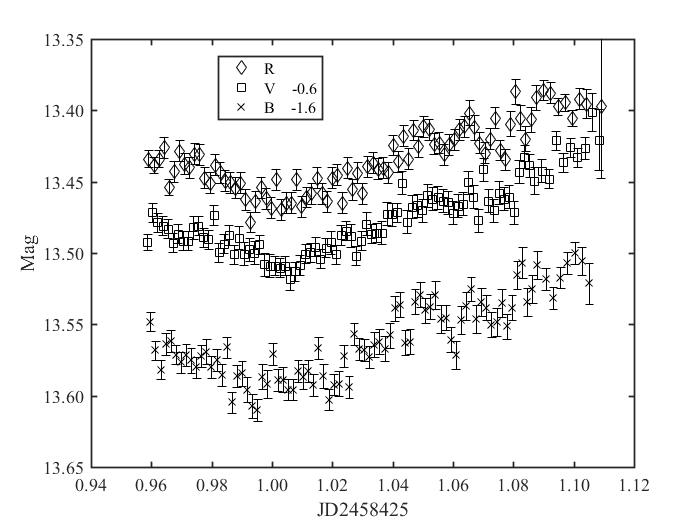}
	\includegraphics[width=0.33\textwidth]{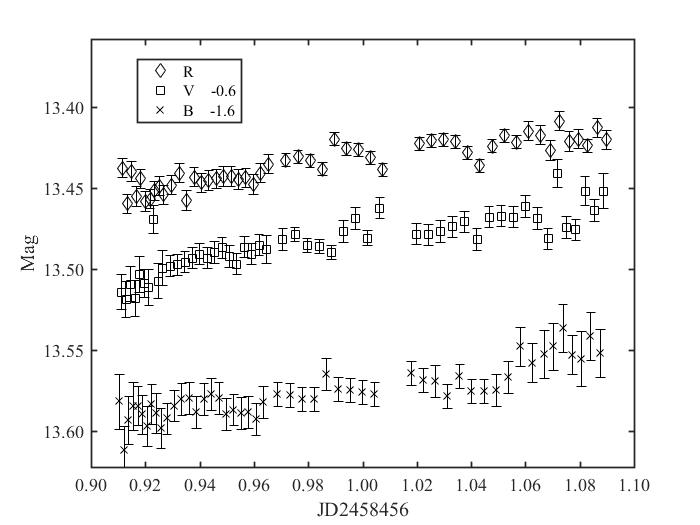}
	\end{tabular}
    \caption{Intraday light curves. For clarity, the B, V, R, and I light curves are shifted. The shifted magnitudes are given in the plots.}
    \label{light curves}
\end{figure*}

\begin{figure}
	\centering
	\includegraphics[width=0.5\columnwidth]{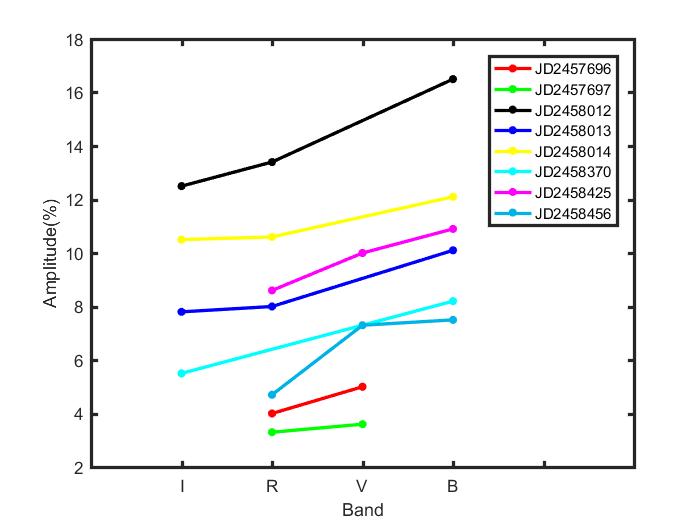}
    \caption{The IDV amplitudes of the variable light curves}
    \label{IDV}
\end{figure}

The intra-night light curves of the object are plotted in Figure \ref{light curves}. \citet{1996A&A...305...42H} developed a method to quantify the IDV amplitude:
\begin{equation}
   A=  \sqrt{(m_{\rm max}-m_{ \rm min})^2-2\sigma^2},
\label{Amplitude equation}
\end{equation}

where $\sigma$ is the measurement error. According to equation $(3)$, the value $A$ will always be larger than 0, however it does not means that the corresponding light curve is variable. The most violent variation happened on JD2458012 (2017 September 15) when the IDV amplitude reached 16.5\% (0.17 mag) in the B band and the variation rate was 0.3 mag/hr. On JD2458014 (2017 September 17) the object reached its brightest state of R = 12.95, while on JD2458420 (2018 October 28) the object was at its faintest state with R = 13.55. The IDV amplitude for each band in each night are included in Table \ref{var}. Figure \ref{IDV} shows the IDV amplitudes of the variable light curves. According to Figure \ref{IDV} and Table \ref{var}, the IDV amplitude is greater in higher energy bands. The IDV amplitude is comparable in a few cases where the differences between IDV amplitude are smaller than 0.5\%. This trend has also been observed by others (e.g. \citealt{1998A&A...332L...1N,1998AJ....115.2244W,1999ApJS..121..131F,2004NuPhS.132..205N,2017MNRAS.469.3588M}). 

\subsection{Variability Detection}
We performed the nested analysis of variance (ANOVA) and enhanced F-test to examine and to quantify the intraday variability \citep{2015AJ....150...44D}. 

In the nested ANOVA analysis, we separated the data points of a certain band on a certain day into groups, with 5 data points in each group. The null hypothesis is that the
deviation of the mean values of differential light curves in each group is zero. The expressions of the degrees of freedom and F-value are shown in the Equation 4 in \citet{2015AJ....150...44D}. 

The enhanced F-test uses several comparison field stars. First, the errors of the field stars are scaled into our target's level. By fitting an exponential curves to our comparison stars we can obtain the relationship between the standard deviation and the mean magnitude of the light curve. This relationship can be used to transfer the error of field stars into the level of BL Lacertae. At last we subtract the mean magnitude from each field star's light curve and stack them together. The F value is the variance of BL Lacertae's curve divided by the variance of the stacked light curve.

\subsection{Variations Result}
The results of the two tests are listed in Table \ref{var}. $\nu1$ and $\nu2$ are degrees of freedom in each test. Column (15) is the IDV amplitude. Column (14) is the Variability. F$_\mathrm{crit}$ is the critical value of F-test at $\alpha$ = 0.05 ($\alpha$ is significance level). When the light curve's F-value is larger than F$_\mathrm{crit}$ and its P-value is smaller than 0.05, the light curve passes the test. If the light curve passed both test, BL Lacertae will be marked as 'Y'(Yes). 'N'(No) means the light curve did not pass at least one of the tests. Twenty-three light curves on twelve nights are variable according to both tests, while three didn't pass both tests. Fourteen light curves passed one of the tests while their P-value of the other test is larger than 0.05, we can not determine their variability, so we mark them as 'P' (Possible). Small flares can be seen in the $V$ band on 2016 November 4 (JD2457697) and 2016 November 06 (JD2457699) , and all bands on 2018 October 28 (JD2458420). Due to the measurement error, these light curves only passed one of the tests but are still very likely variable.
\begin{table*}
	\centering
	\caption{IDV test results.}
	\label{var}
	\fontsize{6}{5}\selectfont
    \begin{tabular}{ccp{3mm}crrrrcrrrrcr} 
    	\hline
    	Julian Date & Date& Filter & \multicolumn{5}{c}{Enhanced F-test}&\multicolumn{5}{c}{Nested ANOVA}&Variability&Amplitude\\
    	 & ( yyyy mm dd )  & & $F$ &$\nu_{1}$& $\nu_{2}$ & $F_{\rm{crit}}$ &\multicolumn{1}{c}{P} & $F$ &$\nu_{1}$& $\nu_{2}$ & $F_{\rm{crit}}$& \multicolumn{1}{c}{P} & &(\%)\\
    	(1)&(2)&(3)&(4)&(5)&(6)&(7)&(8)&(9)&(10)&(11)&(12)&(13)&(14)&(15)\\
    	\hline
    	2457696 & 2016 11 03 & $V$ & 3.67 & 160 & 800 & 1.21 & $<0.0001$ & 5.10 & 31 & 159& 1.83 & $<0.0001$ & Y & 5.0\\
    	 &  & $R$ & 5.89 & 160 & 800 & 1.21 & $<0.0001$ & 2.46 & 31 & 159 & 1.83 & $<0.0001$ & Y & 4.0\\
    	&&&&&&&&&&&&&\\
    	2457697 & 2016 11 04 & $V$ & 3.02 & 94 & 470 & 1.28 & $<0.0001$ & 1.92 & 17 & 89 & 2.21 & $<0.0001$ & Y&3.6\\
    	 &  & $R$ & 6.52 & 93 & 465 & 1.29 & $<0.0001$ & 3.79 & 17 & 89 & 2.21 & $<0.0001$ & Y & 3.3\\
    	 &&&&&&&&&&&&&\\
    	2457699 & 2016 11 06 & $V$ & 1.08 & 178 & 890 & 1.21 & $<0.0001$ & 2.28 & 18 & 94 & 1.79 & 0.0082 & N&\\
    	 &  & $R$ & 0.11 & 181 & 905 & 1.25 & 0.9999 & 1.51 & 19 & 99 & 1.77 & 0.1044 & N &\\
    	 &&&&&&&&&&&&&\\
    	2457745 & 2016 12 22 & $B$ & 1.89 & 97 & 485 & 1.28 & $<0.0001$ & 1.14 & 18 &94 & 2.21 & 0.3297 & P&5.4\\
    	 &  & $V$ & 2.80 & 97 & 485 & 1.28 & $<0.0001$ & 0.48 & 18 & 94 & 2.21 & 0.9584& P&4.1\\
    	 & & $R$ & 2.06 & 95 & 475 & 1.28 & $<0.0001$ & 1.06 &18 &94 & 2.21 & 0.4118 & P & 7.5\\
    	 &&&&&&&&&&&&&\\
    	2457746 & 2016 12 23 & $B$ & 1.36 & 103 & 515 & 1.27 & 0.0165 &3.08 & 19 & 99 & 2.12 & $<0.0001$ & Y&10.6\\
    	 &  & $V$ & 2.10 & 101 & 505 & 1.27 & $<0.0001$ & 1.29 & 19 & 99 & 2.12 & 0.0686 & P&7.5\\
    	 &  & $R$ & 2.28 & 101 & 505 & 1.27 & $<0.0001$ & 1.41& 19 & 99 & 2.12 & 0.5124 & P&7.4\\
    	 &&&&&&&&&&&&&\\
    	2458012 & 2017 09 15 & $B$ & 20.15 & 232 & 1160 & 1.18 & $<0.0001$ & 8.46 & 45 & 229 & 1.66 & $<0.0001$ & Y & 16.5\\
        & & $R$ & 28.59 & 250 & 1250 & 1.17 & $<0.0001$ & 8.07 & 49 & 249 & 1.63 & $<0.0001$ & Y & 13.4\\
    	 &  & $I$ & 28.69 & 233 & 1165 & 1.18 & $<0.0001$ & 7.43 & 45 & 229 & 1.66 & $<0.0001$ & Y & 12.5\\
    	 &&&&&&&&&&&&&\\
    	2458013 & 2017 09 16 & $B$ & 7.99 & 235 & 1175 & 1.17 & $<0.0001$ & 3.33 & 46 & 234 & 1.66 & $<0.0001$ & Y & 10.1\\
    	 &  & $R$ & 13.87 & 255 & 1275 & 1.17 & $<0.0001$ & 3.61 & 50 & 254 & 1.63 & $<0.0001$ & Y & 8.0\\
    	 &  & $I$ & 6.14 & 232 & 1160 & 1.18 & $<0.0001$ & 3.62 & 45 & 229 & 1.66 & $<0.0001$ & Y & 7.8\\
    	 &&&&&&&&&&&&&\\
    	2458014 & 2017 09 17 & $B$ & 6.46 & 230 & 1150 & 1.18 & $<0.0001$ & 2.45 & 45 & 229 & 1.66 & $<0.0001$ & Y & 12.1\\
    	 &  & $R$ & 12.56 & 236 & 1180 & 1.17 & $<0.0001$ & 2.74 & 46 & 234 & 1.66 & $<0.0001$ & Y & 10.6\\
    	 &  & $I$ & 16.54 & 228 & 1140 & 1.18 & $<0.0001$ & 4.89 & 44 & 224 & 1.68 & $<0.0001$ & Y & 10.5\\
    	 &&&&&&&&&&&&&\\
    	2458369 & 2018 09 07 & $B$ & 1.02 & 31 & 155 & 1.53 & 0.4436 & 1.56  & 5 & 29 & 2.62 & 0.2106 & N&\\
    	          &            & $V$ & 4.44 & 31 & 155 & 1.53 & $<0.0001$ & 2.02  & 5 & 29 & 2.62 & 0.1123 & P&3.3\\
    	          &            & $R$ & 4.76 & 31 & 155 & 1.53 & $<0.0001$ & 6.94 & 5 & 29 & 2.62 & 0.0003 & Y & 3.2\\
    	          &            & $I$ & 4.88 & 31 & 155 & 1.53 & $<0.0001$ & 1.25 & 5 & 29 & 2.62 & 0.3176 & P & 3.3\\
    	         
    	&&&&&&&&&&&&&\\
    	2458370 & 2018 09 08 & $B$ & 8.00 & 52 & 260 & 1.40 & $<0.0001$ & 4.29  & 9 & 49 & 2.12 & 0.0006 & Y & 8.2\\
    	          &            & $V$ & 8.66 & 52 & 260 & 1.40 & $<0.0001$ & 1.96  & 9 & 49 & 2.12 & 0.0706 & P&5.6\\
    	          &            & $R$ & 14.22 & 52 & 260 & 1.40 & $<0.0001$ & 1.59  & 9 & 49 & 2.12 & 0.1523& P &4.9\\
    	          &            & $I$ & 27.30 & 52 & 260 & 1.40 & $<0.0001$ & 3.50 & 9 & 49 & 2.12 & 0.0028& Y & 5.5\\
    	          
    	&&&&&&&&&&&&&\\
    	2458420 & 2018 10 28 & $B$ & 2.37 & 67 & 335 & 1.34 & $<0.0001$ & 1.66  & 12 & 64 & 1.94 & 0.1040 &P & 5.6\\
    	          &            & $V$ & 3.01 & 68 & 340 & 1.34 & $<0.0001$ & 1.49  & 12 & 64 & 1.94 & 0.1597 & P & 4.4 \\
    	          &            & $R$ & 3.81 & 68 & 340 & 1.34 & $<0.0001$ & 0.99 & 12 & 64 & 1.94 & 0.4700 &P & 3.3 \\
    	&&&&&&&&&&&&&\\
    	2458424 & 2018 11 01 & $R$ & 2.64 & 132 & 660 & 1.24 & $<0.0001$ & 1.09 & 25 & 129 & 1.94 & 0.3716 & P & 6.3\\
    	           &  & $I$ & 3.87 & 132 & 660 & 1.24 & $<0.0001$ & 3.93 & 25 & 129 & 1.94 & 0.0008 & Y & 6.4 \\
    	&&&&&&&&&&&&&\\
    	2458425 & 2018 11 02 & $B$ & 4.31 & 84 & 420 & 1.30 & $<0.0001$ & 2.82 & 15 & 79 & 2.31 & 0.0020 & Y&10.9\\
    	 &  & $V$ & 7.12 & 84 & 420 & 1.30 & $<0.0001$ & 4.28 & 15 & 79 & 2.31 & $<0.0001$ & Y&10.0\\
    	 &  & $R$ & 7.28 & 84 & 420 & 1.30 & $<0.0001$ & 2.50 &15 & 79 & 2.31 & 0.0046 & Y & 8.6\\
    	&&&&&&&&&&&&&\\
    	2458456 & 2018 12 03 & $B$ & 1.52 & 51 & 255 & 1.40 & 0.0195    & 6.27  & 9 & 49 & 2.12 & $<0.0001$ & Y & 7.5\\
    	          &            & $V$ & 7.32 & 51 & 255 & 1.40 & $<0.0001$ & 4.99  & 9 & 49 & 2.12 & 0.002& Y & 7.3\\
                  &            & $R$ & 7.73 & 51 & 255 & 1.40 & $<0.0001$ & 2.25 & 9 & 49 & 2.12 & 0.0382 & Y&4.7\\
    	\hline
    \end{tabular}

\end{table*}
\subsection{Inter-band correlation analysis and time lags}
To search for the possible time lags between variations in different bands, we performed cross-correlation analyses. Two cross-correlation methods are used. One is the z-transformed discrete correlation functions (ZDCFs) \citep{1997ASSL..218..163A,2013arXiv1302.1508A}, in which equal population binning and Fisher's z-transform are used to correct several biases of the discrete correlation function of \citet{1988ApJ...333..646E}. A Gaussian Fitting (GF) of the ZDCF points greater than 75\% of the peak value can give an estimate of the time lag and the associated errors. However, Gaussian fitting may underestimate the error (e.g. \citealt{2012AJ....143..108W}). So we only take the results of ZDCF+GF method as a reference. For more reliable estimate of the time lags and errors, we used the interpolated cross-correlation function (ICCF)\citep{1987ApJS...65....1G}. \citet{1998PASP..110..660P,2004ApJ...613..682P} employed a Monte-Carlo (MC) method to calculate the centroid position of the ICCF and its error. Flux-randomization (FR) and random-subset selection (RSS) are applied in each Monte-Carlo realization. Here we performed 5000 Monte-Carlo realizations. The results are listed in Table \ref{lag}. On 2018 December 3 (JD2458456), the Gaussian fitting failed to fit the ZDCF, the results are not listed in the table.

According to our results, no time lag has significance greater than 3$\sigma$, thus we failed to detect any time lags at high significance in our observations.

\begin{figure*}
    \centering
    \begin{tabular}{ccc}
	\includegraphics[width=0.33\columnwidth]{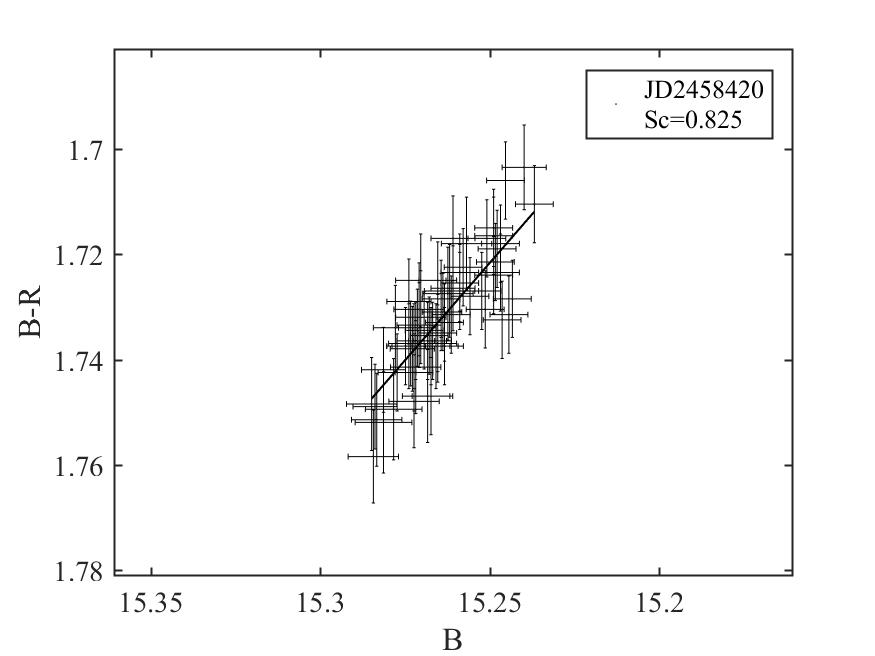}
	\includegraphics[width=0.33\columnwidth]{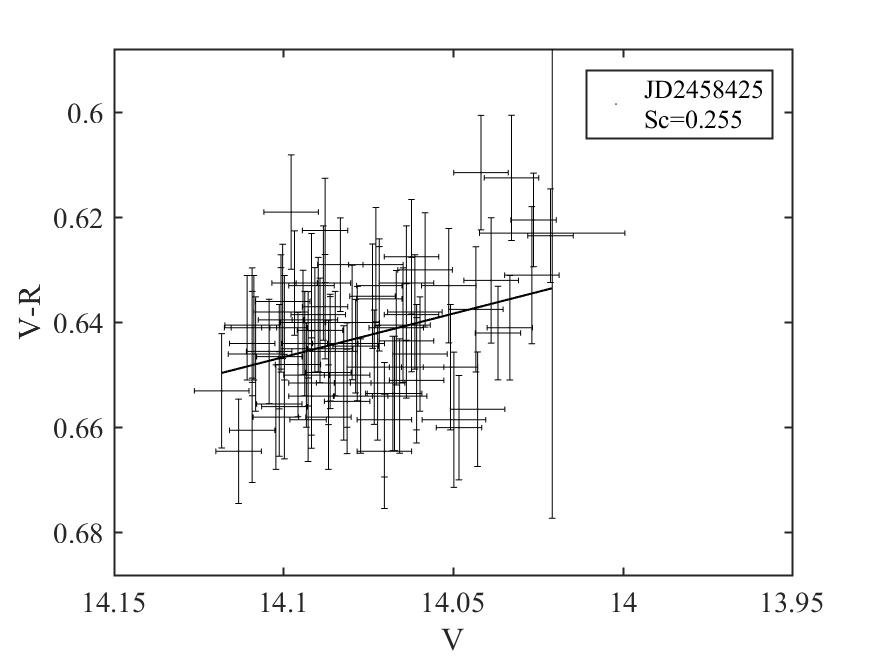}
	\includegraphics[width=0.33\columnwidth]{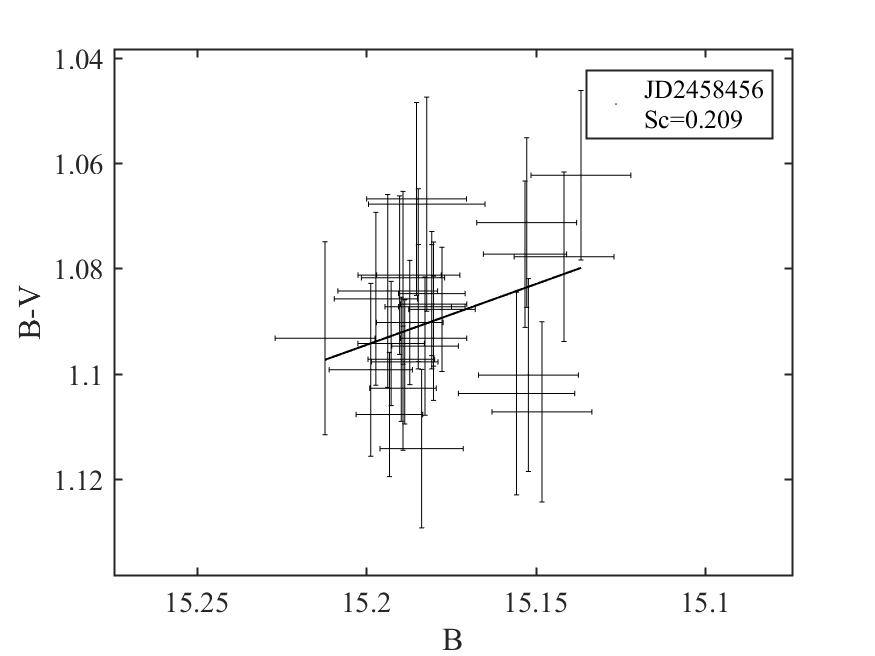}
	\end{tabular}
    \caption{Examples of intranight colour-magnitude diagrams, which illustrate strong, mild and weak correlations, respectively.}
    \label{colorfig}
\end{figure*}

\begin{table*}
	\centering
	\caption{Time lags from correlation analyses}
	\label{lag}
	\begin{tabular}{cclr@{$\pm$}lr@{$\pm$}l} 
		\hline
		Julian Date & Date & Passbands & \multicolumn{2}{c}{ZDCF$-$GF} & \multicolumn{2}{c}{ICCF-FR/RSS}\\
		&( yyyy mm dd )&& \multicolumn{2}{c}{(minutes)} & \multicolumn{2}{c}{(minutes)}\\
		\hline
		2457696 & 2016 11 03 & $R-V$ & $-$28.86 & 9.13 & $-$22.62 & 29.01 \\
		2457697 & 2016 11 03 & $R-V$ & $-$7.07 & 5.29 & $-$5.73 & 6.18\\
		2457699 & 2016 11 06 & $R-V$ & $-$15.59 & 4.49 & $-$5.69 & 5.05 \\
		2457745 & 2016 12 22 & $R-V$ & $-$8.02 & 10.63 & $-$6.03 & 38.42 \\
		          &            & $R-B$ & $-$11.68 & 7.67 & $-$18.56 & 36.13 \\   
		          &            & $V-B$ & $-$3.90 & 1.94 & $-$4.08 & 25.57 \\  
		2457746 & 2016 12 23 & $R-V$ & $-$2.77 & 4.78 & $-$2.14 & 9.12  \\
		          &            & $R-B$ & $+$7.49 & 4.34 & $+$10.67 & 13.58  \\ 
		          &            & $V-B$ & $+$10.40 & 6.24 & $+$12.45 &12.76 \\ 
		2458012 & 2017 09 15 & $R-I$ & $-$7.63 & 3.35 & $-$3.34 & 4.54\\
		          &            & $R-B$ & $-$23.87 & 5.24 & $-$2.61 & 6.18 \\
		          &            & $I-B$ & $-$0.38 & 3.16 & $+$2.22 & 3.51 \\
		2458013 & 2017 09 16 & $R-I$ & $+$4.05 & 1.29 & $-$1.63 & 4.45\\
		          &            & $R-B$ & $+$2.54 & 1.05 & $+$0.13 & 3.93\\
		          &            & $I-B$ & $+$2.00 & 0.74 & $+$2.85 & 2.63\\
		2458014 & 2017 09 17 & $R-I$ & $-$2.25 & 2.92 & $+$5.59 & 6.88\\
		          &            & $R-B$ & $-$9.74 & 2.63 & $-$3.86 & 8.64\\
		          &            & $I-B$ & $-$0.44 & 8.97 & $-$7.13 & 5.92\\
		2458369 & 2018 09 07 & $B-R$ & $-$15.37 & 7.08 & $-$15.11 & 29.98\\
		          &            & $B-V$ & $-$4.68 & 10.73 & $-$3.62 & 32.65\\
		          &            & $B-I$ & $+$11.53 & 5.33 & $+$7.02 & 36.80\\
		2458370 & 2018 09 08 & $B-R$ & $+$2.20 & 4.78 & $+$1.30 & 7.27\\
		          &            & $B-V$ & $+$1.56 & 4.36 & $+$0.41 & 7.47\\
		          &            & $B-I$ & $+$6.15 & 5.5 & $+$3.04 & 6.52\\
		2458420 & 2018 10 28 & $R-V$ & $+$33.49 & 9.74 & $+$20.02 & 91.25\\
		          &            & $R-B$ & $+$19.22 & 6.94 & $+$16.63 & 92.97\\
		          &            & $V-B$ & $-$16.86 & 11.63 & $-$57.09 & 66.31\\
		2458424 & 2018 11 01 & $R-I$ & $-$5.4 & 4.05 & $-$0.58 & 3.21\\
		2458425 & 2018 11 02 & $R-V$ & $-$2.39 & 1.07 & $-$1.70 & 6.34\\
		          &            & $R-B$ & $-$4.47 & 1.16 & $+$1.69 & 6.36\\
		          &            & $V-B$ & $-$2.87 & 1.41 & $-$1.20 & 6.56\\
		2458456 & 2018 12 03 & $R-V$& \multicolumn{2}{c}{} &$+$28.94 & 53.46\\
		          &            & $R-B$ & \multicolumn{2}{c}{}&$-$10.22 & 33.68\\
		          &            & $V-B$ & \multicolumn{2}{c}{}&$+$19.81 & 46.77\\
		\hline
	\end{tabular}
\end{table*}

\subsection{Color Variations}
By calculating the colour indices of $B$ $-$ $R$, $V$ $-$ $R$, $V$ $-$ $I$, and $B$ $-$ $I$, we investigated the color variation with respect to the magnitude of BL Lacertae for each night. A linear fit was made to the data points. The examples of color-magnitude diagrams are shown in Figure \ref{colorfig}. The scales of the horizontal-axis and vertical-axis are fixed as 0.2 and 0.1 mags, respectively in order to make comparison conveniently between panels. 

We calculated the Spearman correlation coefficient (Sc) and the corresponding P value. The results are displayed in Table \ref{color}. The first and second column are the date of observation, the third and fourth column are the passbands of color index, the fifth column is the number of datapoints followed by Sc value and corresponding P value. The last column the the strength of correlation. The bands without variation are not shown in this table because it is pointless to discuss their trend. The numbers of data points of our color-magnitude diagrams are all above 31. The critical value of the correlation coefficient is 0.442 at the significance level of 0.01 and degree of freedom of 31. Most of our Sc value of color-magnitude diagrams are above 0.344 with P value lower than 0.0001, indicating strong correlations. On 2017 September 17 (JD2458014), the Sc values of all the color-magnitude diagram are lower than the critical values with P values larger than 0.01, we take them as no correlations. On 2018 October 28 (JD2458420) the Sc value of B$-$R diagram is 0.820 with the P value significantly smaller than 0.01, indicating a strong BWB color behavior. For the rest, the Sc value ranges from 0.4 to 0.6 with P value smaller than 0.001. In the B and R bands of  2018 December 3 (JD2458456), R and V bands of 2018 December 3 (JD2458456), the light curves are not correlated.

According to \citet{2011PASJ...63..639I}, we do not know if the BWB trend is universal in blazars. There are two types of blazars, BL Lacertae objects and FSRQs. The BWB trend is often observed in BL Lac objects while the RWB behavior is frequently detected in FSRQs (e.g. \citealt{2006A&A...450...39G,2018MNRAS.478.3513Z, 2018RAA....18..150L, 2019AJ....157...95G}). \citet{2002A&A...390..407V,2004A&A...421..103V} argued that the BWB relation was more likely to be detected in short isolated outbursts. The BWB trend of BL Lacertae has been detected in a number of previous research  \citep{1970ApJ...159L..99R,1998A&A...339..382S,2003MmSAI..74..963V,2006MNRAS.366.1337S,2007A&A...470..857P,2011PASJ...63..639I,2015MNRAS.452.4263G,2015A&A...573A..69W,2018ApJS..237...30M}. Our intraday color-magnitude results consist with most of the historical observations.

\begin{table*}
	\centering
	\caption{ The result of Colour-magnitude diagrams.}
	\label{color}
	\begin{tabular}{p{15mm}|cccclccc} 
		\hline
		Julian Date & Date & Color index & Magnitude &\multicolumn{1}{c}{No.} & Sc & \multicolumn{1}{c}{P}  & correlation\\
		&( yyyy mm dd )&&\\
		\hline
		2457696 & 2016 11 03 & $R-V$ & $V$ & 160 & 0.50 & $6.6\times10^{-12}$  &strong \\
		2457697 & 2016 11 04 & $R-V$ & $V$ & 93& 0.43 & $1.3\times10^{-5}$  & strong \\
		2457745 & 2016 12 22 & $R-V$ & $V$ & 95& 0.48 & $6.3\times10^{-7}$  & strong \\
		          &            & $R-B$ & $B$ & 95  & 0.64 & $1.3\times10^{-12}$& strong \\   
		          &            & $V-B$ & $B$ & 97 & 0.64 & $1.5\times10^{-12}$  & strong \\  
		2457746 & 2016 12 23 & $R-V$ & $V$ & 101 & 0.54 & $6.5\times10^{-9}$ & strong \\
		          &            & $R-B$ & $B$ & 101 & 0.66 & $4.6\times10^{-14}$ & strong \\ 
		          &            & $V-B$ & $B$ & 101& 0.54 & $3.0\times10^{-9}$  & strong \\ 
		2458012 & 2017 09 15 & $I-R$ & $R$ & 233 & 0.26 & $6.9\times10^{-5}$ & strong \\
		          &            & $R-B$ & $B$ & 232 & 0.62 & $4.5\times10^{-26}$ & strong \\
		          &            & $I-B$ & $B$ & 232 & 0.71 & $1.0\times10^{-36}$ &strong \\
		2458013 & 2017 09 16 & $I-R$ & $R$ & 232 & 0.36 & $1.2\times10^{-8}$ & strong \\
		          &            & $R-B$ & $B$ & 235 & 0.46 & $7.1\times10^{-14}$ & strong \\
		          &            & $I-B$ & $B$ & 235 & 0.57 & $4.1\times10^{-21}$ & strong \\
		2458014 & 2017 09 17 & $I-R$ & $I$ & 228 & 0.12 & $7.9\times10^{-2}$ & no \\
		          &            & $R-B$ & $B$ & 230 & 0.04 & $5.6\times10^{-1}$ & no \\
		          &            & $I-B$ & $B$ & 228 & 0.04 & $5.7\times10^{-1}$ & no \\
		2458369 & 2018 09 07 & $B-I$ & $B$ & 31 & 0.65 & $1.0\times10^{-4}$ & strong \\
		2458370 & 2018 09 08 & $B-R$ & $B$ & 52 & 0.62 & $9.5\times10^{-7}$ & strong \\
		          &            & $B-I$ & $B$ & 52 & 0.42 & $2.1\times10^{-3}$ & strong \\
		2458420 & 2018 10 28 & $R-V$ & $V$ & 68 & 0.70 & $2.8\times10^{-11}$ & strong \\
		          &            & $R-B$ & $B$ & 67 & 0.82 & $9.5\times10^{-18}$ & strong \\
		          &            & $V-B$ & $B$ & 67 & 0.64 & $4.6\times10^{-9}$ & strong \\
		2458424 & 2018 11 01 & $I-R$ & $R$ & 132 & 0.36 & $1.8\times10^{-6}$ & strong \\
		2458425 & 2018 11 02 & $R-V$ & $V$ & 84 & 0.26 & $1.9\times10^{-2}$ & mild \\
		          &            & $R-B$ & $B$ & 84 & 0.58 & $7.7\times10^{-9}$ & strong \\
		          &            & $V-B$ & $B$ & 84 & 0.49 & $1.8\times10^{-6}$ & strong \\
		2458456 & 2018 12 03 & $R-V$ & $V$ & 51 & 0.49 & $1.8\times10^{-1}$ & no \\
		          &            & $R-B$ & $B$ & 51 & 0.20 & $2.6\times10^{-1}$ & no \\
		          &            & $V-B$ & $B$ & 51 & 0.21 & $3.0\times10^{-4}$ & weak \\
		\hline
	\end{tabular}
\end{table*}

\section{Discussions and Conclusions}

We monitored BL Lacertae in the $B$, $V$, $R$ and $I$ bands for 14 nights during 2016-2018. The object showed IDV in 23 light curves on 12 nights. 

It has been found that the IDV amplitude of BL Lacertae is greater at higher frequencies (e.g. \citealt{2001IAUS..205...82K,2001A&A...369..758F,2003A&A...397..565P,2006MNRAS.373..209H,2015MNRAS.452.4263G,2017MNRAS.469.3588M}). This behavior has also been detected in our observation. \citet{2021APh...12902577B} found similar behavior for S5 0716+714. \citet{2015MNRAS.452.4263G} interpreted this behavior as higher energy electrons accelerated by shock front lose energy faster than low energy electrons through synchrotron radiation. Hence the amount of higher frequency photons produced by these electrons will have a more violent change than the lower frequency photons. This will be observed as IDV amplitude is greater at higher frequencies. The higher energy electron is produced in a thin layer behind the shock front and lower-frequency emission is spread out behind the shock front \citep{1985ApJ...298..114M}, it results in time lags of the peak of the light curve toward lower frequencies. However, this trend is not universal in every observation. The flux of bluer bands is lower than redder bands and has higher errors than redder bands (Figure \ref{light curves} shows the magnitude of B band is larger than R band for BL Lacertae). The error component in equation \ref{Amplitude equation} will reduce the IDV amplitude, so the IDV amplitude of bluer bands with higher errors will be reduced more than the IDV amplitude of redder bands. For example on 2017 September 16th (JD2458013), R and I have comparable IDV amplitudes.

BWB trend was found in our observation. This is consistent with previous (e.g. \citealt{2003MmSAI..74..963V,2015MNRAS.452.4263G,2015A&A...573A..69W,2017MNRAS.469.3588M,2018MNRAS.478.3513Z,2017Galax...5...94G,2018Galax...6....2B,2012A&A...538A.125Z}). We did not subtract the contribution of host galaxy from the total flux since \citet{2002A&A...390..407V} concluded that the color changes are intrinsic property of fast flares and are not related to the host galaxy contribution. \citet{2006MNRAS.373..209H} also argued that the host galaxy and AGN has similar color so the color changes of AGN are not affected by the host galaxy. \citet{2015A&A...573A..69W} did a long-term observation of the BL Lacertae object and argued that the BWB trend is less likely caused by the host galaxy if the color-magnitude diagram shows separate branches. It is believed that the BWB trend is originated from the emission regions of the jet. As the object gets brighter, more relativistic electrons will be accelerated and injected into the emission zone. The high energy photons from the synchrotron mechanism typically emerge sooner and closer to the shock front than the lower energy ones, thus causing color variations and larger variation amplitude at higher frequencies \citep{2002ApJ...564...92C,2004A&A...419...25F}. According to \citet{2020ApJ...888...30F}, the significance of BWB trend might be affected by the strength of variation, the BWB trend caused by the shock will be more significant during a weaker phase of variation and vice versa. 

No time-lag has been detected in our observation. \citet{2012AJ....143..108W} mentioned four key parameters that might determine whether the time-lag can be detected: wavelength separation, variation amplitude, temporal resolution and measurement accuracy. According to Table 1, our temporal resolution (less than 5 minutes) is much smaller than previous detected time lag. Three or four filters were used to observe BL Lacertae on 10 nights which provided us large wavelength separation. Small variation amplitude (JD2457697) and low measurement accuracy (JD2458420, JD2458424) might be the reason why we didn't detect time-lag. In addition, correlation analysis will fail to detect the time lag (if there is any) between featureless or monotonically brightening or darkening light curves, such as those on JD 2457696, 2458012, and 2458014. 

In conclusion, BL Lacertae showed IDV in 23 light curves on 12 nights among our 14 nights observation during 2016-2018. We found the IDV amplitude of BL Lacertae is greater at higher frequencies. In addition, BL Lacertae shows BWB trend on most of the nights. Finally, no time-lag has been detected in our observation.

\section*{Acknowledgements}

This work has been supported by the National Natural Science Foundation of China grants 11973017.
  
\bibliographystyle{raa}
\bibliography{RAA-2021-0173.R1}

\begin{thebibliography}{61}
\providecommand\natexlab[1]{#1}
\providecommand\JournalTitle[1]{#1}

\bibitem[{Abdo} {et~al.}(2010)]{2010ApJ...716...30A}
{Abdo}, A.~A., {Ackermann}, M., {Agudo}, I., {et~al.} 2010, \apj, 716, 30

\bibitem[{Agarwal} \& {Gupta}(2015)]{2015MNRAS.450..541A}
{Agarwal}, A., \& {Gupta}, A.~C. 2015, \mnras, 450, 541

\bibitem[{Alexander}(1997)]{1997ASSL..218..163A}
{Alexander}, T. 1997, {Is AGN Variability Correlated with Other AGN Properties?
  ZDCF Analysis of Small Samples of Sparse Light Curves}, ed. D.~{Maoz},
  A.~{Sternberg}, \& E.~M. {Leibowitz}, Vol. 218, Astronomical Time Series, ed.
  D.~{Maoz}, A.~{Sternberg}, \& E.~M. {Leibowitz}, Vol. 218, 163

\bibitem[{Alexander}(2013)]{2013arXiv1302.1508A}
{Alexander}, T. 2013, arXiv e-prints, arXiv:1302.1508

\bibitem[{Bach} {et~al.}(2006)]{2006A&A...456..105B}
{Bach}, U., {Villata}, M., {Raiteri}, C.~M., {et~al.} 2006, \aap, 456, 105

\bibitem[{Bhatta} \& {Webb}(2018)]{2018Galax...6....2B}
{Bhatta}, G., \& {Webb}, J. 2018, Galaxies, 6, 2

\bibitem[{B{\"o}ttcher} {et~al.}(2003)]{2003ApJ...596..847B}
{B{\"o}ttcher}, M., {Marscher}, A.~P., {Ravasio}, M., {et~al.} 2003, \apj, 596,
  847

\bibitem[{Butuzova}(2021)]{2021APh...12902577B}
{Butuzova}, M.~S. 2021, Astroparticle Physics, 129, 102577

\bibitem[{Carini} {et~al.}(1992)]{1992AJ....104...15C}
{Carini}, M.~T., {Miller}, H.~R., {Noble}, J.~C., \& {Goodrich}, B.~D. 1992,
  \aj, 104, 15

\bibitem[{Chiang} \& {B{\"o}ttcher}(2002)]{2002ApJ...564...92C}
{Chiang}, J., \& {B{\"o}ttcher}, M. 2002, \apj, 564, 92

\bibitem[{Ciprini} {et~al.}(2004)]{2004MNRAS.348.1379C}
{Ciprini}, S., {Tosti}, G., {Ter{\"a}sranta}, H., \& {Aller}, H.~D. 2004,
  \mnras, 348, 1379

\bibitem[{de Diego} {et~al.}(2015)]{2015AJ....150...44D}
{de Diego}, J.~A., {Polednikova}, J., {Bongiovanni}, A., {et~al.} 2015, \aj,
  150, 44

\bibitem[{Edelson} \& {Krolik}(1988)]{1988ApJ...333..646E}
{Edelson}, R.~A., \& {Krolik}, J.~H. 1988, \apj, 333, 646

\bibitem[{Epstein} {et~al.}(1972)]{1972ApJ...178L..51E}
{Epstein}, E.~E., {Fogarty}, W.~G., {Hackney}, K.~R., {et~al.} 1972, \apjl,
  178, L51

\bibitem[{Fan} \& {Lin}(1999)]{1999ApJS..121..131F}
{Fan}, J.~H., \& {Lin}, R.~G. 1999, \apjs, 121, 131

\bibitem[{Fan} {et~al.}(2001)]{2001A&A...369..758F}
{Fan}, J.~H., {Qian}, B.~C., \& {Tao}, J. 2001, \aap, 369, 758

\bibitem[{Feng} {et~al.}(2020)]{2020ApJ...888...30F}
{Feng}, H.-C., {Liu}, H.~T., {Bai}, J.~M., {et~al.} 2020, \apj, 888, 30

\bibitem[{Fiorucci} {et~al.}(2004)]{2004A&A...419...25F}
{Fiorucci}, M., {Ciprini}, S., \& {Tosti}, G. 2004, \aap, 419, 25

\bibitem[{Fossati} {et~al.}(1998)]{1998MNRAS.299..433F}
{Fossati}, G., {Maraschi}, L., {Celotti}, A., {Comastri}, A., \& {Ghisellini},
  G. 1998, \mnras, 299, 433

\bibitem[{Gaskell} \& {Peterson}(1987)]{1987ApJS...65....1G}
{Gaskell}, C.~M., \& {Peterson}, B.~M. 1987, \apjs, 65, 1

\bibitem[{Gaur} {et~al.}(2017)]{2017Galax...5...94G}
{Gaur}, H., {Gupta}, A., {Bachev}, R., {et~al.} 2017, Galaxies, 5, 94

\bibitem[{Gaur} {et~al.}(2015)]{2015MNRAS.452.4263G}
{Gaur}, H., {Gupta}, A.~C., {Bachev}, R., {et~al.} 2015, \mnras, 452, 4263

\bibitem[{Goyal} {et~al.}(2013)]{2013JApA...34..273G}
{Goyal}, A., {Mhaskey}, M., {Gopal-Krishna}, {et~al.} 2013, Journal of
  Astrophysics and Astronomy, 34, 273

\bibitem[{Gu} {et~al.}(2006)]{2006A&A...450...39G}
{Gu}, M.~F., {Lee}, C.~U., {Pak}, S., {Yim}, H.~S., \& {Fletcher}, A.~B. 2006,
  \aap, 450, 39

\bibitem[{Gupta} \& {Joshi}(2005)]{2005A&A...440..855G}
{Gupta}, A.~C., \& {Joshi}, U.~C. 2005, \aap, 440, 855

\bibitem[{Gupta} {et~al.}(2019)]{2019AJ....157...95G}
{Gupta}, A.~C., {Gaur}, H., {Wiita}, P.~J., {et~al.} 2019, \aj, 157, 95

\bibitem[{Heidt} \& {Wagner}(1996)]{1996A&A...305...42H}
{Heidt}, J., \& {Wagner}, S.~J. 1996, \aap, 305, 42

\bibitem[{Hu} {et~al.}(2006)]{2006MNRAS.373..209H}
{Hu}, S.~M., {Wu}, J.~H., {Zhao}, G., \& {Zhou}, X. 2006, \mnras, 373, 209

\bibitem[{Ikejiri} {et~al.}(2011)]{2011PASJ...63..639I}
{Ikejiri}, Y., {Uemura}, M., {Sasada}, M., {et~al.} 2011, \pasj, 63, 639

\bibitem[{Kurtanidze} {et~al.}(2001)]{2001IAUS..205...82K}
{Kurtanidze}, O.~M., {Richter}, G.~M., \& {Nikolashvili}, M.~G. 2001, in
  Galaxies and their Constituents at the Highest Angular Resolutions, ed. R.~T.
  {Schilizzi}, Vol. 205, 82

\bibitem[{Li} {et~al.}(2018)]{2018RAA....18..150L}
{Li}, X.-P., {Luo}, Y.-H., {Yang}, H.-T., {et~al.} 2018, Research in Astronomy
  and Astrophysics, 18, 150

\bibitem[{Marscher}(2014)]{2014ApJ...780...87M}
{Marscher}, A.~P. 2014, \apj, 780, 87

\bibitem[{Marscher} \& {Gear}(1985)]{1985ApJ...298..114M}
{Marscher}, A.~P., \& {Gear}, W.~K. 1985, \apj, 298, 114

\bibitem[{Meng} {et~al.}(2017)]{2017MNRAS.469.3588M}
{Meng}, N., {Wu}, J., {Webb}, J.~R., {Zhang}, X., \& {Dai}, Y. 2017, \mnras,
  469, 3588

\bibitem[{Meng} {et~al.}(2018)]{2018ApJS..237...30M}
{Meng}, N., {Zhang}, X., {Wu}, J., {Ma}, J., \& {Zhou}, X. 2018, \apjs, 237, 30

\bibitem[{Miller} \& {Hawley}(1977)]{1977ApJ...212L..47M}
{Miller}, J.~S., \& {Hawley}, S.~A. 1977, \apjl, 212, L47

\bibitem[{Nesci} {et~al.}(1998)]{1998A&A...332L...1N}
{Nesci}, R., {Maesano}, M., {Massaro}, E., {et~al.} 1998, \aap, 332, L1

\bibitem[{Nikolashvili} \& {Kurtanidze}(2004)]{2004NuPhS.132..205N}
{Nikolashvili}, M.~G., \& {Kurtanidze}, O.~M. 2004, Nuclear Physics B
  Proceedings Supplements, 132, 205

\bibitem[{Padovani} \& {Giommi}(1995)]{1995MNRAS.277.1477P}
{Padovani}, P., \& {Giommi}, P. 1995, \mnras, 277, 1477

\bibitem[{Papadakis} {et~al.}(2003)]{2003A&A...397..565P}
{Papadakis}, I.~E., {Boumis}, P., {Samaritakis}, V., \& {Papamastorakis}, J.
  2003, \aap, 397, 565

\bibitem[{Papadakis} {et~al.}(2007)]{2007A&A...470..857P}
{Papadakis}, I.~E., {Villata}, M., \& {Raiteri}, C.~M. 2007, \aap, 470, 857

\bibitem[{Peterson} {et~al.}(1998)]{1998PASP..110..660P}
{Peterson}, B.~M., {Wanders}, I., {Horne}, K., {et~al.} 1998, \pasp, 110, 660

\bibitem[{Peterson} {et~al.}(2004)]{2004ApJ...613..682P}
{Peterson}, B.~M., {Ferrarese}, L., {Gilbert}, K.~M., {et~al.} 2004, \apj, 613,
  682

\bibitem[{Racine}(1970)]{1970ApJ...159L..99R}
{Racine}, R. 1970, \apjl, 159, L99

\bibitem[{Raiteri} {et~al.}(2009)]{2009A&A...507..769R}
{Raiteri}, C.~M., {Villata}, M., {Capetti}, A., {et~al.} 2009, \aap, 507, 769

\bibitem[{Sadun} {et~al.}(2020)]{2020Galax...8...11S}
{Sadun}, A.~C., {Asadi-Zeydabadi}, M., {Hindman}, L., \& {Moody}, J.~W. 2020,
  Galaxies, 8, 11

\bibitem[{Sandrinelli} {et~al.}(2018)]{2018A&A...615A.118S}
{Sandrinelli}, A., {Covino}, S., {Treves}, A., {et~al.} 2018, \aap, 615, A118

\bibitem[{Scarpa} {et~al.}(2000)]{2000ApJ...544..258S}
{Scarpa}, R., {Urry}, C.~M., {Padovani}, P., {Calzetti}, D., \& {O'Dowd}, M.
  2000, \apj, 544, 258

\bibitem[{Smith} {et~al.}(1985)]{1985AJ.....90.1184S}
{Smith}, P.~S., {Balonek}, T.~J., {Heckert}, P.~A., {Elston}, R., \& {Schmidt},
  G.~D. 1985, \aj, 90, 1184

\bibitem[{Speziali} \& {Natali}(1998)]{1998A&A...339..382S}
{Speziali}, R., \& {Natali}, G. 1998, \aap, 339, 382

\bibitem[{Stalin} {et~al.}(2006)]{2006MNRAS.366.1337S}
{Stalin}, C.~S., {Gopal-Krishna}, {Sagar}, R., {et~al.} 2006, \mnras, 366, 1337

\bibitem[{Vagnetti} \& {Trevese}(2003)]{2003MmSAI..74..963V}
{Vagnetti}, F., \& {Trevese}, D. 2003, \memsai, 74, 963

\bibitem[{Villata} {et~al.}(2002)]{2002A&A...390..407V}
{Villata}, M., {Raiteri}, C.~M., {Kurtanidze}, O.~M., {et~al.} 2002, \aap, 390,
  407

\bibitem[{Villata} {et~al.}(2004{\natexlab{a}})]{2004A&A...421..103V}
{Villata}, M., {Raiteri}, C.~M., {Kurtanidze}, O.~M., {et~al.}
  2004{\natexlab{a}}, \aap, 421, 103

\bibitem[{Villata} {et~al.}(2004{\natexlab{b}})]{2004A&A...424..497V}
{Villata}, M., {Raiteri}, C.~M., {Aller}, H.~D., {et~al.} 2004{\natexlab{b}},
  \aap, 424, 497

\bibitem[{Wagner} \& {Witzel}(1995)]{1995ARA&A..33..163W}
{Wagner}, S.~J., \& {Witzel}, A. 1995, \araa, 33, 163

\bibitem[{Webb} {et~al.}(1998)]{1998AJ....115.2244W}
{Webb}, J.~R., {Freedman}, I., {Howard}, E., {et~al.} 1998, \aj, 115, 2244

\bibitem[{Wierzcholska} {et~al.}(2015)]{2015A&A...573A..69W}
{Wierzcholska}, A., {Ostrowski}, M., {Stawarz}, {\L}., {Wagner}, S., \&
  {Hauser}, M. 2015, \aap, 573, A69

\bibitem[{Wu} {et~al.}(2012)]{2012AJ....143..108W}
{Wu}, J., {B{\"o}ttcher}, M., {Zhou}, X., {et~al.} 2012, \aj, 143, 108

\bibitem[{Zhai} \& {Wei}(2012)]{2012A&A...538A.125Z}
{Zhai}, M., \& {Wei}, J.~Y. 2012, \aap, 538, A125

\bibitem[{Zhang} {et~al.}(2018)]{2018MNRAS.478.3513Z}
{Zhang}, X., {Wu}, J., \& {Meng}, N. 2018, \mnras, 478, 3513

\end{thebibliography}

\end{document}